\documentclass[a4paper]{article}
\usepackage[dvips]{graphicx}
\usepackage[dvips]{color}

\title{Track detector DEVIS for the double beta decay investigation}

\author{V.\,Artemiev, V.\,Belov, E.\,Brakhman, A.\,Karelin, V.\,Kirichenko, \\
O.\,Kozodaeva, A.\,Kuchenkov, V.\,Lubimov, A.\,Mitin, T.\,Tsvetkova, \\
O.\,Zeldovich}

\date{23 November 2004}

\begin{document}
\maketitle

\begin{center}
Institute for Theoretical and Experimental Physics
\end{center}

\begin{abstract}
Detector DEVIS is TPC in the magnetic field. It is dedicated to the 
investigation of the double-beta decay of Xe. Setup sensitivity was 
estimated in the series of measurements with Xe with natural isotopes 
composition. Detector allows measuring two-neutrino double-beta decay 
of $^{136}$Xe with half-life less than $3 \cdot 10^{20}$ years.
\end{abstract}

\section{Introduction}

Results obtained in the solar, atmospheric and reactor experiments are 
interpreted as the evidence of the neutrino oscillations and the existence 
of a small ($m_{\nu} > 0.045$~eV) neutrino mass. But absolute values of neutrino 
masses and their hierarchy cannot be determined solely from oscillation 
experiments. Experimental searches for the neutrinoless double beta decay 
enable to check up the law of lepton charge conservation, to define the 
neutrino mass nature (Dirac or Majorana) and to investigate the absolute 
scale of neutrino masses~\cite{1,2}. Evidence of the observation of $^{76}$Ge 
$2\beta0\nu$-decay~\cite{3} gave rise to a great discussion. Thus appears the 
necessity to check this result and its interpretation with other isotopes and 
with higher sensitivity.

A lot of new projects of the experiments with few tons isotopes are under 
consideration and development at present, for example GENIUS~\cite{4} with $^{76}$Ge 
and EXO~\cite{5} with $^{136}$Xe. The goal of the future experiments is to reach of 
the sensitivity $T \sim 10^{27-28}$ years and to a neutrino mass of the order 
$m_{\nu} < 0.05$~eV. However, for unequivocal definition on a neutrino mass (or 
limit on mass) the knowledge of a nuclear matrix element is necessary. 
There are many publications, devoted to theoretical calculations of the 
nuclear matrix elements both for $2\beta2\nu$ and $2\beta0\nu$ decay modes. 
Results differ 
more than 2 times. Therefore, experimental observations of $2\beta2\nu$-decay for 
different isotopes with maximal accuracy are extremely important. Possible 
test of the theoretical calculations has been proposed in~\cite{6}. Where 
it was stated that final test of the theoretical calculations can be 
accomplished if $2\beta0\nu$-decay of three (or more) isotopes will be observed.

$^{76}$Ge $2\beta2\nu$-decay was measured in the few experiments~\cite{7,8}, 
though some 
small disagreement exists. $^{136}$Xe $2\beta2\nu$-decay is not observed up to 
now, so existing limits for half life time greatly differ, 
from $T \sim 10^{20}$~y to $T \sim 10^{22}$~y~\cite{9,10,11,12,13}.

Large track detector DEVIS (DEtector VISualizing) is the Time Projection 
Chamber (TPC) in the magnetic field and it is operating at ITEP~\cite{14}. 
The first goal of the experiment is the investigation of $^{136}$Xe 
$2\beta2\nu$-decay.

\section{Setup description}

\begin{figure}\begin{center}
\includegraphics{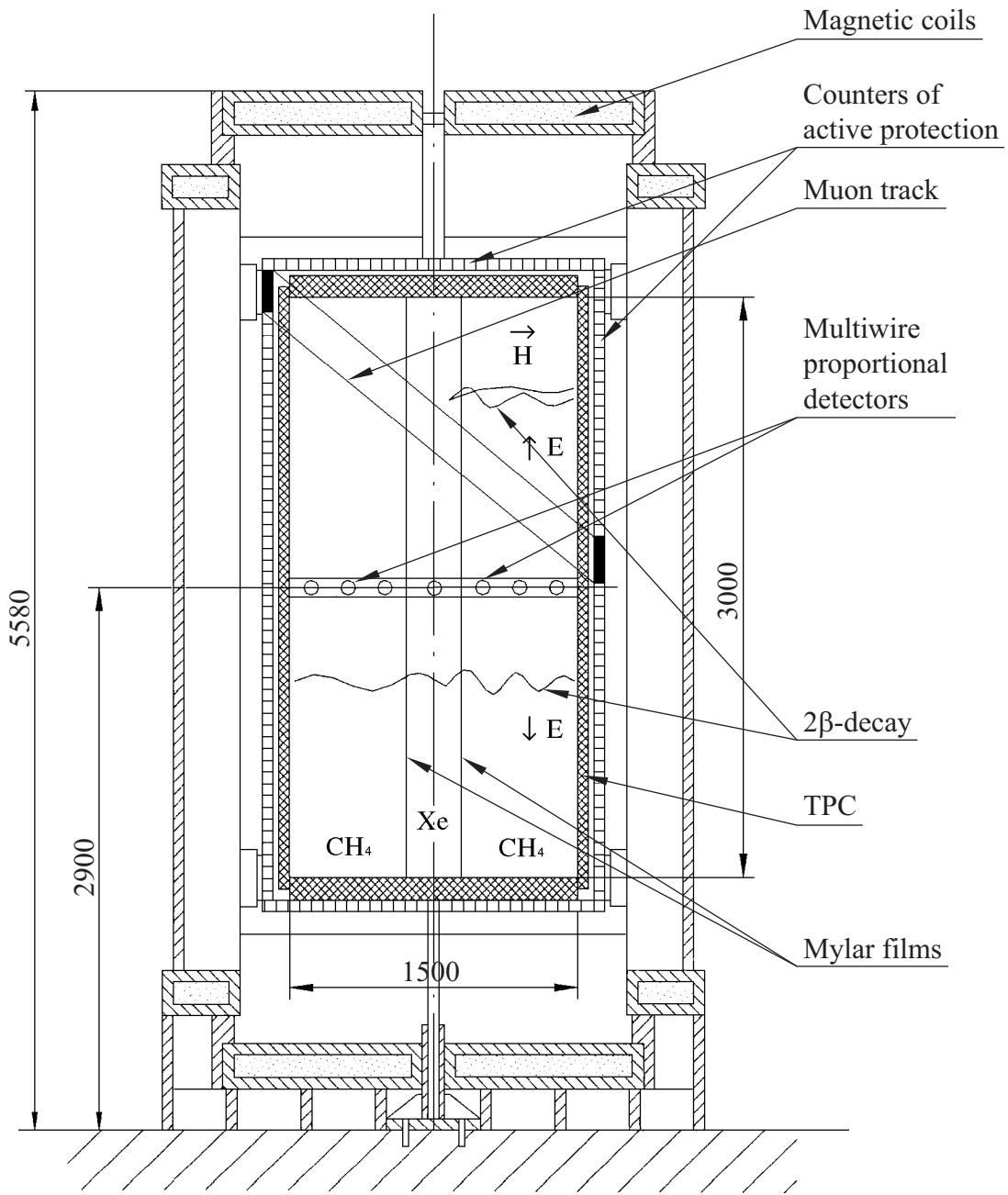} 
\caption{The schematic view of the setup}\label{pic_setup}
\end{center}\end{figure}

Schematic view of the setup one can see on the Fig.~\ref{pic_setup}. TPC is placed at 
the magnet with diameter $5.5$ m. Magnetic field can be changed in the 
range $0.5-1.0$ kilogauss with homogeneity not worse than 2\%.

Source of $2\beta$-decay is $^{136}$Xe, located in the central part of TPC, 
separated by thin mylar films (50~mkm) out of two adjacent parts, 
filled by methane. Detector is operating at atmospheric pressure. 
TPC detector has two drift sections, upper and lower (high voltage drift 
electrodes are located on the upper and lower plates). Uniformly 
distributed potential from 35~kV to 0~kV is fixed on the all-lateral 
surfaces (including mylar films) using special electrodes, creating the 
homogeneous electric drift field. Ionization from passed charge particles 
in the gas is drifted in direction of an electric field to the proportional 
chambers, arranged in the center of TPC.

Three multiwire chambers (for each gas volume) have two rows of sensitive 
wires registering the events in the upper and lower drift sections. Signals 
from two sensitive wires (the upper and the lower) are summarized on the 
input of preamplifier. Detector has $2\cdot(56+56+17)$ sensitive wires. Additional 
readout from stripes (cathode wire electrodes formed $45^{\circ}$ related to the 
sensitive wires) is made for methane volumes~\cite{15}. 

The coordinate measurements are carried out with TDC (Time Digital Converter) 
with number of the sensitive wire being coordinate along the magnetic field 
while the drift time being the coordinate along the electric field. The 
trajectory of the electron in magnetic field has a spiral form. Therefore 
in the basic projection (number of sensitive wire -- drift time) track 
looks like a sinusoid, and in the additional projection (number of cathode 
strip -- drift time) like a cycloid. The momentum and motion direction are 
defined by fitting of the trajectory.

\section{Electronics}

\begin{figure}\begin{center}
\includegraphics[width=\textwidth]{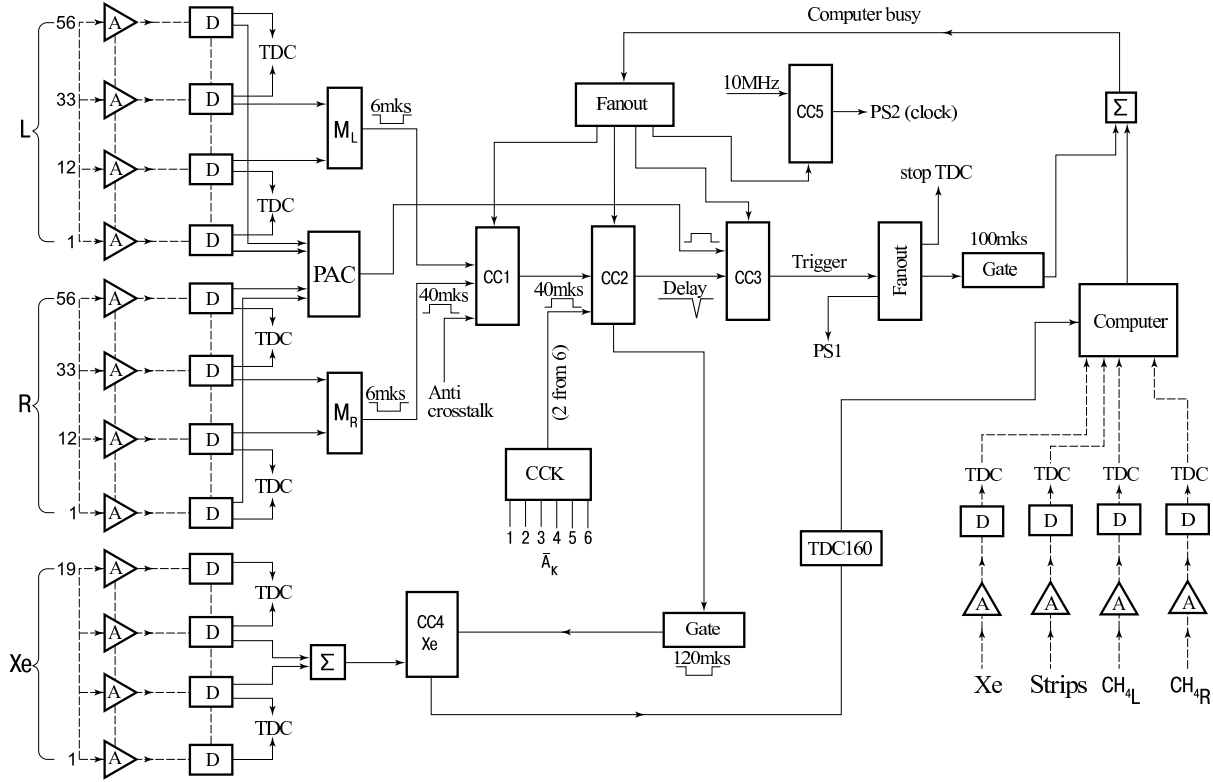}
\caption{Scheme of the trigger and recording electronics}\label{pic_el}
\end{center}\end{figure}

Block scheme of the trigger and recording electronics one can see on 
Fig.~\ref{pic_el}. Signals from preamplifiers of sensitive wires and cathode strips [A] 
with 13~m cables go to discriminators [D] with threshold 30~mV (15~mV) 
with standard NIM ($0.7$~V) output. The discriminator can work in three 
modes: duration above threshold, formation of the duration, start and end 
of the signal from the wire or strip. We mainly used duration above the 
threshold mode. Thus duration on an output of the discriminator corresponds 
to the duration of a signal exceeding the threshold. Dead time of the 
device was equal to 60~ns. Discriminators have two outputs. Each wire and 
strip was connected to TDC using twisted pairs and the time of each hit 
from wires or strips was fixed. Other outputs from discriminators of 
sensitive wires were used to form trigger. Parameters of TDC (full time 
of TDC circle -- 80~mks, one bin -- 20~ns) were chosen to minimize the 
errors in spatial resolution due to electronics, compared to errors by 
multiple scattering and diffusion. Since it is unknown when an event 
happens the setup worked in mode ``RUN''. The clock generator was started 
by computer, and stopped by trigger. Trigger started information recording 
from TDC to computer. The signal ``Fast Dump'' [FD] cleared the memory of 
all TDC if trigger did not occurred during 320~mks, 4 cycles of TDC. FD 
was blocked if it was coincided with trigger, so we had not losses due 
to FD. On the first stage we used 320 TDC blocks~\cite{16}.

The basic part of trigger electronic consists of two trigger processors 
M$_{L}$ and M$_{R}$ connected to left and right volumes of TPC. The trigger signal 
from processor arises, when more than 20 sensitive wires from 32 wires 
on the inputs of M$_{L,R}$, coincide in the 6 mks gate. This time corresponds 
to the size of the sinusoid track and drift velocity in the CH$_4$ volume. 
The signals from M$_{L}$ and M$_{R}$ coincide at the scheme [CC1]. At the first 
stage we used as a trigger coincidences between 4 fixed wires (2 wires 
from both left and right CH$_4$ volumes). Use of the processors increased the 
trigger efficiency from large drift distance because the register 
efficiency of separate sensitive wires is equal to 85--92\% depended 
on drift pass.

A signal from another processor [PAC] arising for events with many tracks 
in methane volumes included as anti-coincidence to the basic trigger [CC3]. 
Usage of this processor allowed skipping directly in the trigger a part of 
events with many tracks, thus we reduced the time of data read-out from 
TDC and increased an alive operating time of setup up to 85\%. Signals 
of counters of the active protection CCK from charge cosmic radiation 
were also included in anti-coincidences to basic trigger [CC2]. The 
description of active protection is resulted in part 5.

At an initial stage we selected events with 2 electrons, directed in 
the different sides. The rate of triggers was 45/s.

\section{Gas supply system}

\begin{figure}\begin{center}
\includegraphics[width=\textwidth]{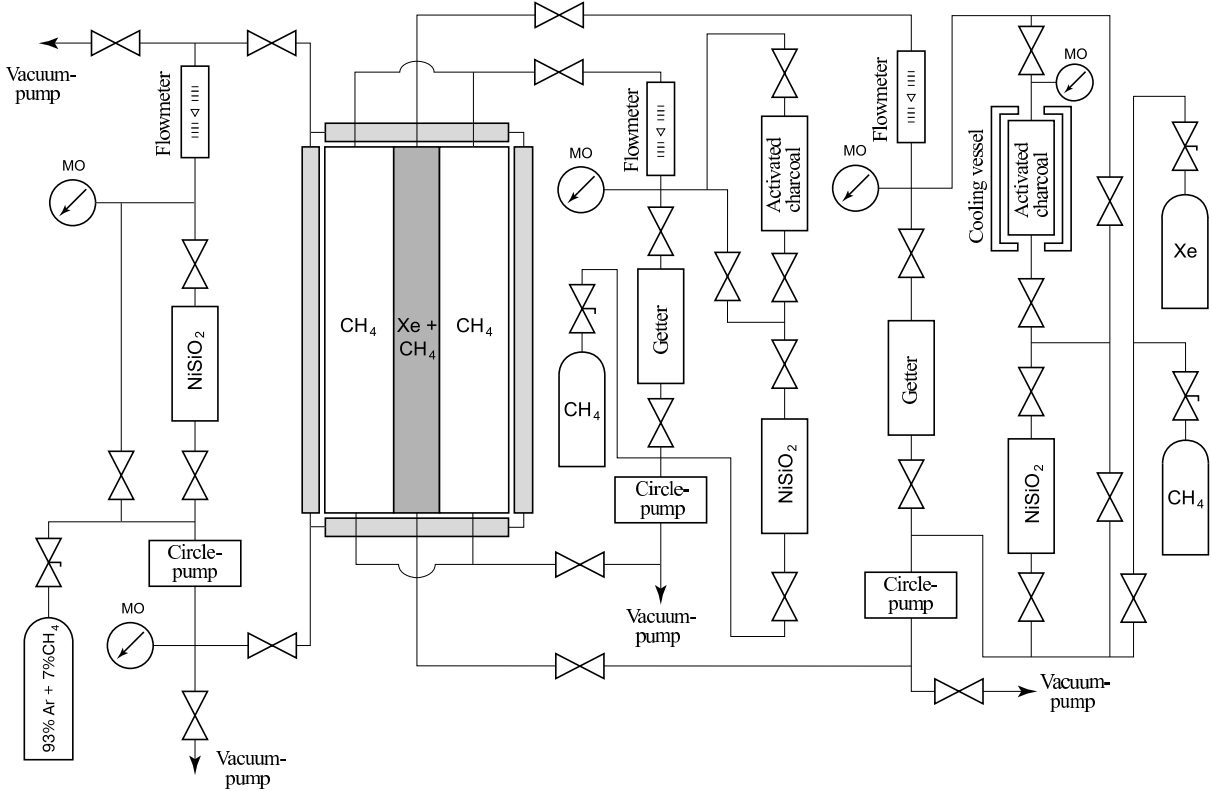}
\caption{Gas supply system}\label{pic_gaz}
\end{center}\end{figure}

As TPC detector consists of three separate not pumped out volumes 
operating at atmospheric pressure filling with working gases and 
maintenance of necessary purity of gases appeared difficult enough. 
Gas supply system consists of 3 separate cycles (see Fig.~\ref{pic_gaz}):
\begin{enumerate}
\item Methane cycle --- for two large side volumes, in sum about 10 m$^3$,
\item Xenon cycle --- for central target volume of 2 m$^3$,
\item Subsystem for the proportional counters of active protection from 
cosmic radiation. 
\end{enumerate}

Two methane volumes were filled by gradual replacing of air with methane, 
at the first stage by pumping-out with submission of methane, then by 
flowing up of necessary quantity of volumes. 

Filling of central target part with mixture Xe\,+\,CH$_4$ was essentially more 
difficult problem, because Xe (even with the natural composition of 
isotopes) is a very expensive gas so flowing up becomes impossible. The 
technique with the chemical reaction described in~\cite{14} was used. First, 
the central volume was filled by CO$_2$. Then CO$_2$ gas was absorbed in 
special reactor with alkali (NaOH) with a compensation of the pressure 
decrease by (Xe\,+\,CH$_4$) mixture. It was necessary to use the admixture 
of methane to increase the drift velocity as such in pure Xe at our 
drift field V$_{dr} = 0.233$~kV/cm was too small. The gas composition was 
controlled in all volumes by chromatograph AXT-002-01 (Russian) with 
accuracy $\sim$~2\%.

After filling the volumes the gas purification is provided in the closed 
cycles. To have high efficiency for large (L $\sim 1.0-1.5$~m) drift distances 
deep purification of the working gases from electronegative impurities 
(O$_2$ and H$_2$O vapor) is necessary up to a level better than 10 ppm. 
Reactors with Ni/SiO$_2$ adsorbent used at the early stage posses fine 
purification properties, but should be used at high temperature 
$\sim 100^{\circ}$C. 
At such temperature the small part of methane dissociates and hydrogen 
appears in the detector volumes what could noticeably change the drift 
velocity. But the main lack of this adsorbent is that it contains the 
big admixture of $^{238}$U and $^{232}$Th. So the isotopes $^{220}$Rn 
and $^{222}$Rn 
appear in the detector due to diffusion from a reactor. The decay of 
their daughter isotopes $^{214}$Bi and $^{208}$Tl could imitate the events of 
double beta decay.

We investigated various getters from the point of view of their cleaning 
properties, capacity on O$_2$ (it was necessary to purify large volumes 
of the gas $10-12$~m$^3$) and the diffusion of Rn isotopes. A getter of the 
active metals alloy was developed specially for us at the Institute of 
Chemistry and Technology of Organic Compound. Two reactors for methane 
and xenon volumes operating at room temperature were prepared. Using of 
new reactors allowed to reduce Rn concentration more than in 50 times. 
Calibrating measurements are described below in Section 10.

Gas electronegative impurities were controlled using amplitude 
measurements with the radioactive source $^{55}$Fe located in windows in 
lateral TPC wall on drift distances of 24 and 129 cm~\cite{14}, and by the 
trigger rate of setup during measurements.

\section{Active protection from charged cosmic particles}

TPC is surrounded by the gas proportional counters of active protection 
from cosmic radiation on all sides. Counters were made of aluminum pipe 
square cross-section $6\times6$~cm$^2$ with 50 mkm central wire. The length of 
counters varies from 1.5 up to 3.5 m. Four in fold joins them in groups 
with one amplifier per fold. Total active surface of 6 panels of counters 
is $\sim 40$~m$^2$. The counters are pumped and filled by CH$_4$/Ar mixture and 
works 3 months without purification, then gas changes. The counters 
efficiency checked with the aid of the radioactive source $^{109}$Cd was not 
worse than 92\%. All 6 panels are included in the coincedence scheme CCK 
(see Fig.~\ref{pic_el}). It gives out a veto signal veto on full time of drift in methane 
in case of signals from any two panels after cosmic particle passage 
through TPC. In addition, the separate segments of protection register 
$\mu$-mesons crossed Xe-volume also giving a veto signal in the trigger 
and registering in TDC. Such including gave the additional suppression 
of events concerned with cosmic particles. Accidental coincidences are 
equal 4\% and are controlled regularly.

\section{Determination of the drift velocity}

To know the drift velocity in methane volume is necessary to calculate 
the electron kinetic energy using a form of the trajectory. The knowledge 
of drift velocity in Xe volume is necessary, because the time difference 
in CH$_4$ and Xe volumes defines the coordinate on drift distance and also 
for restoration of event vertex.

Measurements of the drift velocities and checks of its dependence on 
drift length are carried out using the cosmic trigger. We look for cosmic 
particles (for example, coincidence of the counters pointed on Fig.~\ref{pic_setup}) 
crossing the volume of TPC. The delay time of each sensitive wire depends 
on drift distance at known cosmic trigger. The centers of the delay time 
distributions for all sensitive wires approximate with a straight line 
well. The inclination of a straight line gives drift velocity as the 
distances between wires are fixed. To test this measurement we also use 
the time distributions for the central wires in CH$_4$, because we know the 
drift distance at fixed trigger (with not so good accuracy). The drift 
velocity in CH$_4$ coincides for both cases and equals W$_{CH_4} = 4.0$~cm/sec. 
It weakly depends on small admixture of nitrogen in volume (accumulated 
from the atmosphere) and was measured regularly during the runs. Changes 
were not more than 10\%. The drift velocity in Xe is determined from time 
distribution for central wire in Xe volume in case of the cosmic trigger 
and ranges within $1.8-0.9$~cm/sec with Xe concentration. 

\begin{figure}\begin{center}
\includegraphics{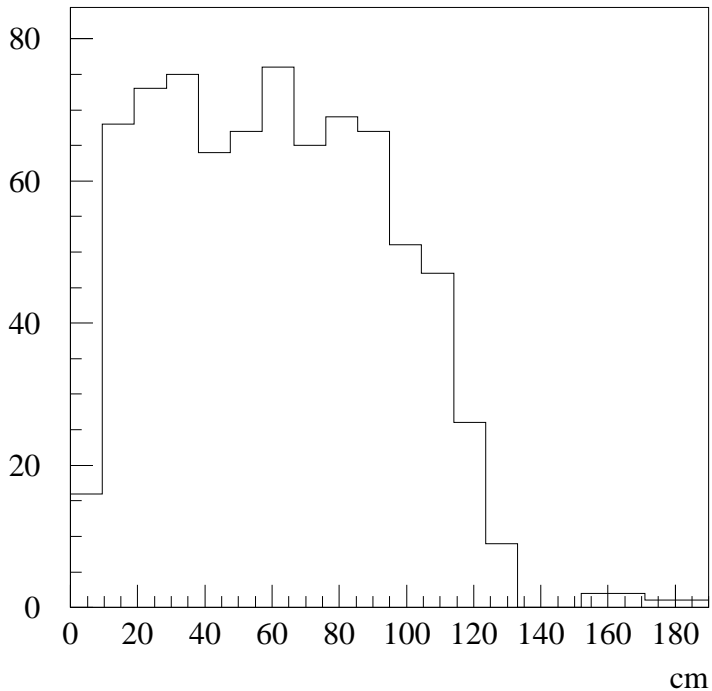} 
\caption{Example of events distribution along drift}\label{pic_dmexe}
\end{center}\end{figure}

The difference in drift times in CH$_4$ and Xe volumes determines the 
absolute event coordinate along an electric drift field 
\begin{displaymath}
L[cm] = (T_{Xe} - T_{CH_4})\cdot(w_{CH_4}\cdot{}w_{Xe})/(w_{Xe}-w_{CH_4})
\end{displaymath}
and is fixed on TDC. At the correct work of the detector the times 
(coordinates) should be distributed uniformly. Distribution of this 
difference between drift times in CH$_4$ and Xe volumes is resulted on 
Fig.~\ref{pic_dmexe}. Such distributions are checked few times in a day and allow 
watching the TPC condition. One can see that small efficiency fall 
takes place along the drift.

\section{Expositions with Xe and test measurements}

A few measurement series were made using different Xe concentration 
with the natural isotope composition for the optimal quantity of Xe 
determination. TPC showed good stability with 50\% concentration of Xe 
and small (4\%) admixture of iC$_4$H$_{10}$ in order to cancel the ``excitation'' 
of Xe wires. On the other hand Monte-Carlo simulations of the TPC 
efficiency for $2\beta$-decay showed the weak dependence on Xe concentration. 
So our measurements were made with the following gas composition: 
Xe~(50\%), CH$_4$~(46\%), iC$_4$H$_{10}$~(4\%). Such concentration corresponds 
to 5 kg of Xe in central part TPC. Drift velocity in the central volume 
decreased up to $0.9$~cm/mks. So to determine the coordinate along the 
drift pass (measuring of the drift time difference in CH$_4$ and Xe volumes) 
special block TDC with total time of 160 mks was used.

During 1000 hours of the exposition broken into two intervals of 515 
and 485 hours because of failure with a high-voltage cable (10~kV supply 
of transformers for the sources of the stabilized current for a magnet) 
about $10^{10}$ triggers were recorded.

Additional expositions with a radioactive source $^{207}$Bi for determination 
of the energy resolution and checking of the efficiency on-line 
programs were also made.

In other runs with enlarged Rn isotopes concentration in the central 
volume of the detector (reactors Ni/SiO$_2$ used) we checked up the quality 
of the new reactors and correctness of work both off-line and modeling 
programs (see Section 10) simultaneously.

Reduction of the Rn concentration by a factor of 50 due to use of reactors 
on the basis of an alloy of metals resulted in the number of double events 
reducing only $6-8$ times. At this level it became clear that an additional 
source of events with two electrons was the cosmic particles crossing the 
dead zones in Xe volume. Active protection from charged cosmic was 
reinforced with few new fold cosmic counters. It decreased quantity of 
double events by $1.5$ times.

The large statistics of events from lateral covers without a direction 
selection has been recorded for the determination of the efficiency of 
separate sensitive wires in CH$_4$ and Xe volumes depended on drift 
distances. The efficiency appeared to be high enough, except for nearest 
to mylar films and to lateral covers wires. These results were used in 
Monte-Carlo calculations.

\section{Selection of \boldmath{$2\beta$}-decay candidate events}

\begin{figure}\begin{center}
\includegraphics{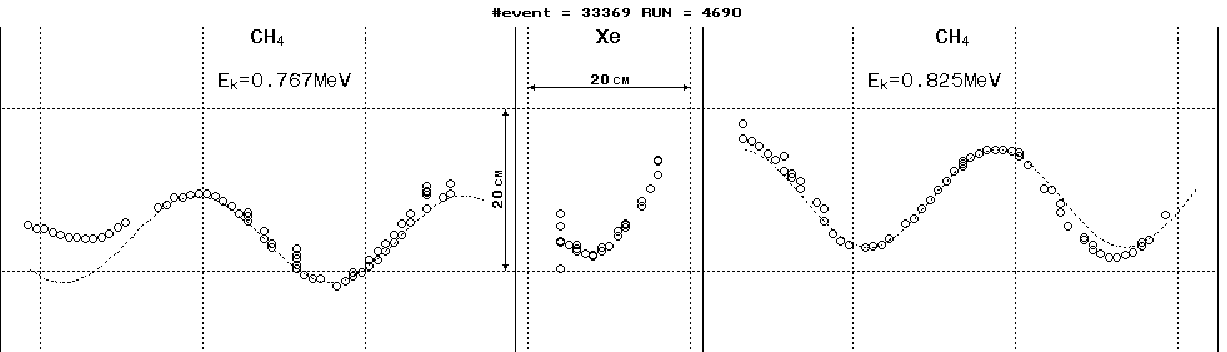}\\ 
\emph{Number of sensitive wire and drift time coordinates}
\includegraphics{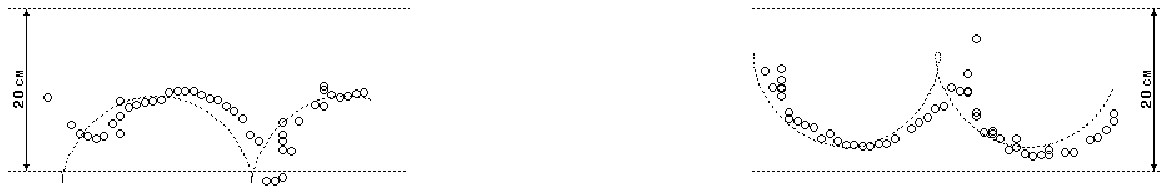}\\ 
\emph{Number of cathode strip and drift time coordinates}
\caption{Example of $2\beta$-event}\label{pic_2b_ev}
\end{center}\end{figure}

Events with electrons escaped from Xe in the two different sides in the 
right and left volumes filled with CH$_4$ are selected. Such selection 
twice reduces efficiency of the detector to $2\beta$-decay, but considerably 
simplifies procedure of events distinguishing as it is not required to 
disentangle electron tracks escaped in one side.

Projection of a track in number of a sensitive wire and drift time 
coordinates is sinusoid, while in number cathodic strip and drift time 
coordinates is cycloid as cathodic strips are located under $45^{\circ}$ angle 
to sensitive wires. The loop or narrow part of cycloid is directed in 
specific direction related to drift depended on a motion direction of 
the electron. So one can select the electrons escaped from Xe volume 
or from lateral covers. Events candidates in double-beta decay had two 
electrons from Xe volumes.

The on-line program uses two simple algorithms for events 
distinguishing. First one for selection of single tracks is restriction 
on number of hits N in each of methane volumes $30 <$~N~$< 100$ on 
56 sensitive wires. Second one is fast algorithm for determination 
of an electrons direction. In a time window related the trigger it 
is determined as $T_{av} = (T_{max} + T_{min})/2$, where $T_{av}$ is average time, 
$T_{max}$, $T_{min}$ -- maximal and minimal drift times for each of methane volumes. 
Then for each methane volume in this time window the numbers of cathode 
strip hits above and below this average drift time are calculated. If 
the number of strip hits above $T_{av}$ exceeds number of strip hits below 
$T_{av}$ it is considered that the narrow part of cycloid (loop) is directed 
downwards to drift that corresponds to one electron direction and in the 
other case to the opposite direction. Such simple algorithms have reduced 
the detector rate from 45 event/s after the electronic trigger up to 2.2 
event/s. The factor of background suppression at this stage is equal 
to 20. The example of $2\beta$-event is presented on Fig.~\ref{pic_2b_ev}.

After on-line program's procession the information is stored on a 
disk and the further off-line analysis of events is proceeded. Further 
this rate has been additionally reduced by $\sim~30$\% by removal of the events 
with an additional track in methane volume (3-rd track), by more careful 
fitting and by removal of events with large energy cosmic particles 
(straight tracks).

Residuary events passes full fit in both number of a sensitive wire and drift 
time coordinates and in number of cathode strip and drift time additional 
coordinates. It should be noted that fitting programs used only tracks 
in methane volumes. In the basic coordinates a projection of electrons 
tracks in each methane volume is fitted by function
\begin{displaymath}
z = A + \rho_L \cdot \sin{\theta} \cdot \sin{\big(\frac{x}{\rho_L \cdot \cos{\theta}} + 
	\varphi \big)}, 
\end{displaymath}
with four parameters, where x is coordinate along which number of a 
sensitive wire varies and z is coordinate along drift. In this case $\theta$ is 
the angle between a momentum of the electron and a magnetic field, $\rho_L$ is 
Larmor radius, A is the mean line of a sinusoid and $\varphi$ is a phase of a 
sinusoid. It is supposed that the momentum and corresponding Larmor 
radius of electron in methane volume changes weakly. Losses in 60~cm of 
CH$_4$ gas make tens keV, and momentum of electron from the double beta 
decay is in a range from 100~keV up to 2~MeV. To take into account 
multiple scattering in methane two breaks on a trajectory are supposed 
during the fitting with Larmor radius being the same but angles $\theta$ and 
$\varphi$ changed. This is especially essential for low energies when scattering 
grows. With one break the number of fitting parameters grows up to 8, 
with two breaks --- up to 12 for 56 points of the data in each of methane 
volume. Software package of minimization MINUIT and software package 
HBOOK from library of programs for processing of physical measurements 
from CERN~\cite{17} were used in fit. 

Since cathode strips are directed relative to sensitive wires under $45^{\circ}$ 
the coordinate along which varies number of a cathodic strip is 
$y_{st}=(x+y)/\sqrt{2}$, where 
x is earlier coordinate along which number of a sensitive wire varied, 
and y is coordinate along this wire. Taking into account that projections 
of a electron trajectory are sinusoids in coordinates z, x and y, x are 
displaced on an angle $\pm\pi/2$ depending on a direction of electron 
movement, we obtain within a constant
\begin{displaymath}
y_{st} = \Big( x + \rho_L \cdot \sin{\theta} \cdot \sin{\big(
	\frac{x}{\rho_L \cdot \cos{\theta}} + 
	\varphi \pm \frac{\pi}{2}\big)}\Big)\Bigm/ \sqrt{2}, 
\end{displaymath}
the equation of two cycloids with loops directed upwards or downwards 
depending on a sign prior $\pi/2$. A constant responsible for shift of 
all cycloid along the $y_{st}$ axis is obtained at fitting events in 
coordinates number of a strip, coordinate along the drift.

\begin{figure}\begin{center}
\includegraphics[width=\textwidth]{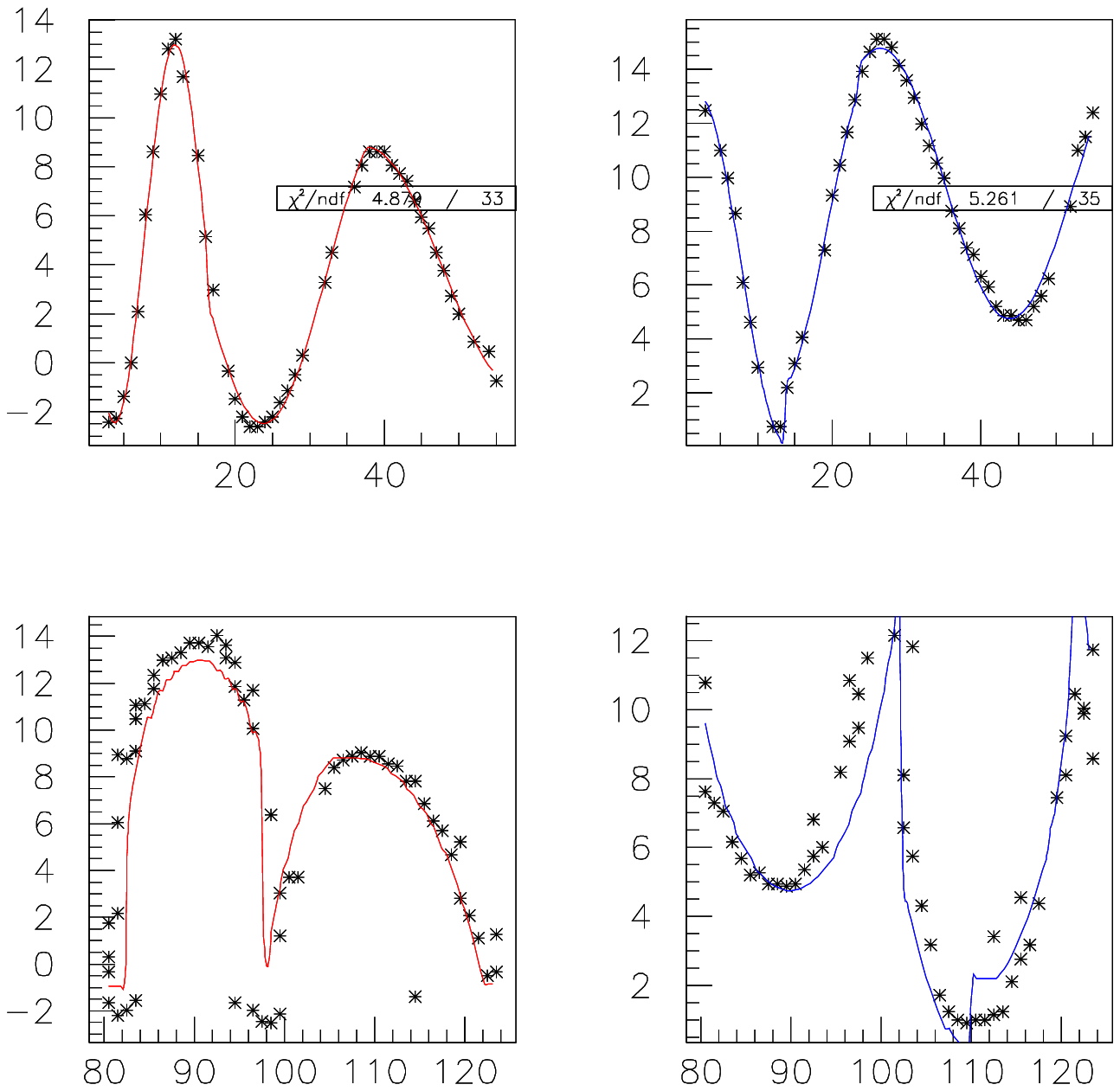}
\caption{Experimental points and fit of $2\beta$-event}\label{pic_fit}
\end{center}\end{figure}

From two hypotheses of a direction of electron the one with the best $\chi^2$ 
is chosen. On Fig.~\ref{pic_fit} experimental points and fit for the candidate 
of $2\beta$-decay event when two electrons escaped from Xe to the left and 
right methane volumes are shown. Below is presented only one of the 
cycloids corresponding to the correct for the candidate in $2\beta$ event 
electron direction and on the right side the loop of a cycloid is 
directed upwards while at the left is directed downwards. Projections 
of a trajectories of the each electron in number of a sensitive wire 
and drift time coordinates (and number of a cathode strip and drift time 
coordinates) are fitted separately. Spatial precision along drift at 
sensitive wires and cathodic strips differs. The signal amplitude from a 
cathodic strip is 6 times less than that from a sensitive wire and on one 
cathode strip there can be added signals from several wires, this leaded 
to shift of the time.

As a result of work of the fitting program the quantity of events had 
decreased from $2.2$~events/s up to $40$~events/hour, i.e. the factor of suppression 
of a background at this stage (it is mainly electrons from lateral covers) 
is~$\sim$~200. Then events are reviewed visually analyzing information from 
both CH$_4$ and Xe volumes.

For determination efficiency of the fitting program the additional runs 
with the increased concentration of $^{222}$Rn in Xe volume were used where 
events with two electrons with a vertex in Xe arose from decay of 
$^{214}$Bi --- daughter nucleus from $^{222}$Rn chain. All of the recorded data 
were viewed without work of fitting programs and after their work. 
Efficiency of the fitting and direction definitions programs is~$\sim$~70\% 
for events with two electrons (see Section 10).

\section{Calculation of the efficiency by a Monte Carlo simulation}

The basic tool of the software for Monte Carlo simulation development 
is package GEANT (versions 3.21) from CERN library~\cite{17}.

On the basis of this package the program model of the detector was 
built. It includes the exact geometry of the detector, the full trigger 
and the emulation of the on-line events selection. Simulation results 
in a file formated completely identically to the data received in real 
experiment. Then this file is analyzed in the same way as the experimental 
data. 

Crossing sensitive wires and strips of the detector forms in 
perpendicular to drift plane a matrix of virtual cells, such that in 
each of them there is a part of one wire and one strip. According to 
the borders of these cells the track of the charged particle breaks 
into slices. For each of them the drift distance is recalculated to 
drift time taking in account diffusion, displacement in a magnetic field 
and efficiency of wires. Received times are written in the corresponding 
channels of wires and strips. The real hardware restrictions are taken 
into account when final times are obtained and transferred to the further 
processing. Direct viewing of the events generated in the simulation has 
not revealed appreciable differences from real ones.

\begin{table}\begin{center}
\parbox[t]{0.6\textwidth}{\caption{Efficiency (\%) and product of efficiency and Xe mass (arbitrary units) 
for $2\beta2\nu$ and $2\beta0\nu$ decay}\label{tab_1}}
\begin{tabular}{|l|c|c|c|c|} \hline
\rule{0pt}{2.5ex}Xe, \%  & \hspace{5mm}20\hspace{5mm} & \hspace{5mm}40\hspace{5mm} & 
\hspace{5mm}50\hspace{5mm} & \hspace{5mm}60\hspace{5mm} \\ \hline
\rule{0pt}{2.5ex}$\varepsilon(2\nu)$         & 1.50 & 0.98 & 0.79  & 0.64  \\ \hline
\rule{0pt}{2.5ex}$\varepsilon(2\nu)\cdot{}m$ & 0.30 & 0.39 & 0.395 & 0.384 \\ \hline
\rule{0pt}{2.5ex}$\varepsilon(0\nu)$         & 12.3 & 10.8 & 9.9   & 9.3   \\ \hline
\rule{0pt}{2.5ex}$\varepsilon(0\nu)\cdot{}m$ & 2.46 & 4.32 & 4.95  & 5.58  \\ \hline
\end{tabular}
\end{center}\end{table}

Efficiency of the detector has been determined using simulation depending 
on Xe concentration. Product of efficiency and number of Xe atoms is the 
value, which is used to obtain the half-life. It shows weak dependence 
from Xe concentration. The results of modeling are listed in the Table 1.

Efficiency for $2\beta2\nu$-decay at chosen 50\% Xe concentration 
$\varepsilon=0.79\%$, while for $2\beta0\nu$-decay $\varepsilon=9.9\%$ 
at work with both drift sections. Small 
efficiency to $2\beta2\nu$-decay is connected with too soft electrons spectrum 
and small probability for electrons to escape from Xe volume and to 
give the trigger.

\section{Detector calibration}

A separate measurements were made to check existing programs modeling 
correctness and to determine efficiency of on-line programs. We used 
a radioactive source $^{207}$Bi, which conversion electrons have energy 
$0.481$ and $0.987$~MeV and lay at energy region of the electrons of $^{136}$Xe 
double beta decay. The intensity of a radioactive source $^{207}$Bi was 
calibrated with the aid of NaI(Tl) detector. The only change in selection 
of events during the work with a radioactive source consisted in changing 
selection by a direction to search events with electrons moving from a 
source through the detector. To reduce background it was used the fact 
that position of a radioactive source is known. Events in the window 
on the strips corresponding to position of a source with a correct value 
of the time difference of signals in CH$_4$ and Xe volume (`a methane-xenon 
difference') were chosen. Measurements with the detector were carried out 
for two positions of the source: near drift (drift distance of 24~cm) and 
far drift (distance of 129~cm).

Efficiency of registration differed~$\sim$~1\% from the modelling data for near 
drift and~$\sim$~10\% for far drift. Such discrepancy can be explained by 
`cancellation' of a source by a background (double tracks) and for far 
drift the additional contribution is brought with distortions of a drift 
field at the large drift distances.

To calibrate the detector and to check all procedures of event selection 
and Monte-Carlo simulation of detector work the additional run~$\sim$~17 hours 
of live time was made. In this run Xe cleaning was carried out by 
Ni/SiO$_2$ reactor resulting in $^{222}$Rn diffusion from the reactor to TPC 
that produced two-electron events with a vertex in Xe volume due to 
$^{214}$Bi decay --- a daughter nucleus of $^{222}$Rn chain.

Quantity of Rn atoms in Xe was determined by measurement of the delayed 
$(\beta-\alpha)$ coincidences from decay of a cascade $^{214}$Bi $\rightarrow$ $^{214}$Po 
with characteristic time $T_{1/2} = 164$~mks (a half-life of $^{214}$Po). 
Since sensitive wires worked in a proportional mode, $\alpha$-particle with 
energy~$\sim$~8~MeV was selected by the amplitude. We used special TDC, 
operating in the mode ``START'' with full time $1.6$ ms. TDC was started by 
electron escaped from Xe volume in one side --- $\beta$ and stopped 
by a signal with the large amplitude in Xe --- $\alpha$. Distribution 
on the arrival time of a signal $\alpha$ answered to a half-life $^{214}$Po. 
Measurement of such coincidences after cleaning Xe volume by Ni/SiO$_2$ 
reactor during $1.5$ hours gave number of coincidences 
$N_{\beta-\alpha} = (527 \pm 32)$/h and when used the reactor on the basis 
of active alloys $N_{\beta-\alpha} = (10.7 \pm 4.3)$/h. Measurements were 
carried out during $1.29$ and $9.25$~hours. Reduction of the $N_{\beta-\alpha}$ 
number with time corresponded to a 
half-life time of $^{222}$Rn $T_{1/2} = 3.8$~days. The similar method was used 
to determine the Rn concentration in CH$_4$ volume. 

Efficiency to $\beta-\alpha$ coincidences was calculated by Monte Carlo 
method using full geometry and imitation of work of programs and was 
$\varepsilon = 4.0$\%. Taking into account the efficiency concentration 
of Rn in Xe volume 
was equal to $N(\mathrm{Rn})/N(\mathrm{Xe+CH_4}) = 3\cdot10^{-20}$ working with getter Ni/SiO$_2$ 
and $0.6\cdot10^{-21}$ using new getter from the active metal alloy.

Efficiency for $2\beta$-events from $^{214}$Bi decay has been calculated too, 
assuming that after decay of daughters Rn ions $^{214}$Bi~\cite{18} was formed 
and fixed on charged films separating gas volumes. Full enough cascade 
of decays $^{214}$Bi~\cite{19} has been used to simulate the events with two 
electrons that can mimicry the $2\beta$-decay. These events appeared as 
$\beta$-decay of $^{214}$Bi with the subsequent emission of the second electron 
due to internal conversion or a birth of Compton or M\"{o}ller electron 
inside Xe-volume. Both electrons escaped to CH$_4$-volumes and gave the 
trigger.

In 17 hours exposition 109 $2\beta$-events were found. Having concentration 
of $^{222}$Rn determined from the delayed $\beta-\alpha$ coincidences 
in view of registration efficiency of events from $^{214}$Bi one could 
expect 119 $2\beta$-events. Observed and expected number of events coincided 
sufficiently well, that gave some confidence of the correctness of all 
selection procedures.

\section{Visual selection of events}

After off-line processing resulted files were viewed visually. Selection 
was carried out by the following criteria. 
\begin{enumerate}
\item Events with a wrong direction of the electron (from a lateral cover) 
were thrown out. A great bulk made such events. The fitting program usually 
made mistakes due to presence of the additional hits (sparks) in wires and 
in strips, and due to the distortion of the tracks as a result of scattering 
and distortion of the electric drift field near the mylar films. 
Strengthening of selection criteria led to the further loss of efficiency.
\item Events with additional tracks in any of methane volumes both in top and 
bottom were thrown out. Events of (1\,+\,2) types were~$\sim$~36/h.
\item Events with two or more tracks in Xe-volume and events when the track 
in Xe was absent (~$\sim$~1/h) were thrown out also.
\item Events with a seen track of a cosmic particle in any of volumes were 
thrown out (~$\sim$~3/h).
\end{enumerate}

Double viewing of the whole material was made. The efficiency of viewing 
was close to 100\%. After visual viewing there were 954 events. In addition 
45 events have been removed from the two-electron events sample with small 
quantity of the worked wires. Events with actuated wires in Xe volume 
N$_{Xe} < 5$ out of total 17 sensitive wires were thrown out. In CH$_4$ volumes 
actuation of N$_{CH_4} > 14$ wires out of 30 nearest to Xe was required. 
Efficiency has not decreased practically, but events with severely 
different number of actuated wires in L and R volumes have been removed, 
that most likely was connected to accidental coincidence in drift of the 
left and right tracks.

909 events recorded during 1000 hours remained for the analysis.

\begin{figure}[htbp]\begin{center}
\includegraphics[width=\textwidth]{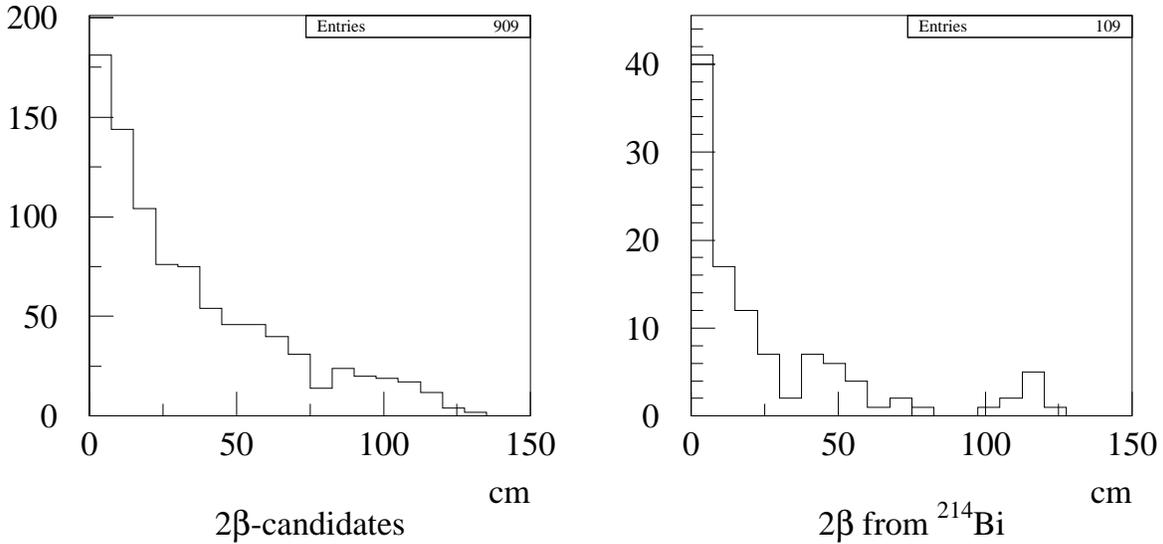}\\
\caption{Distribution on drift distance for $2\beta$-events}\label{pic_dmexe2}
\end{center}\end{figure}

\begin{figure}[htbp] \begin{center}
\includegraphics[width=\textwidth]{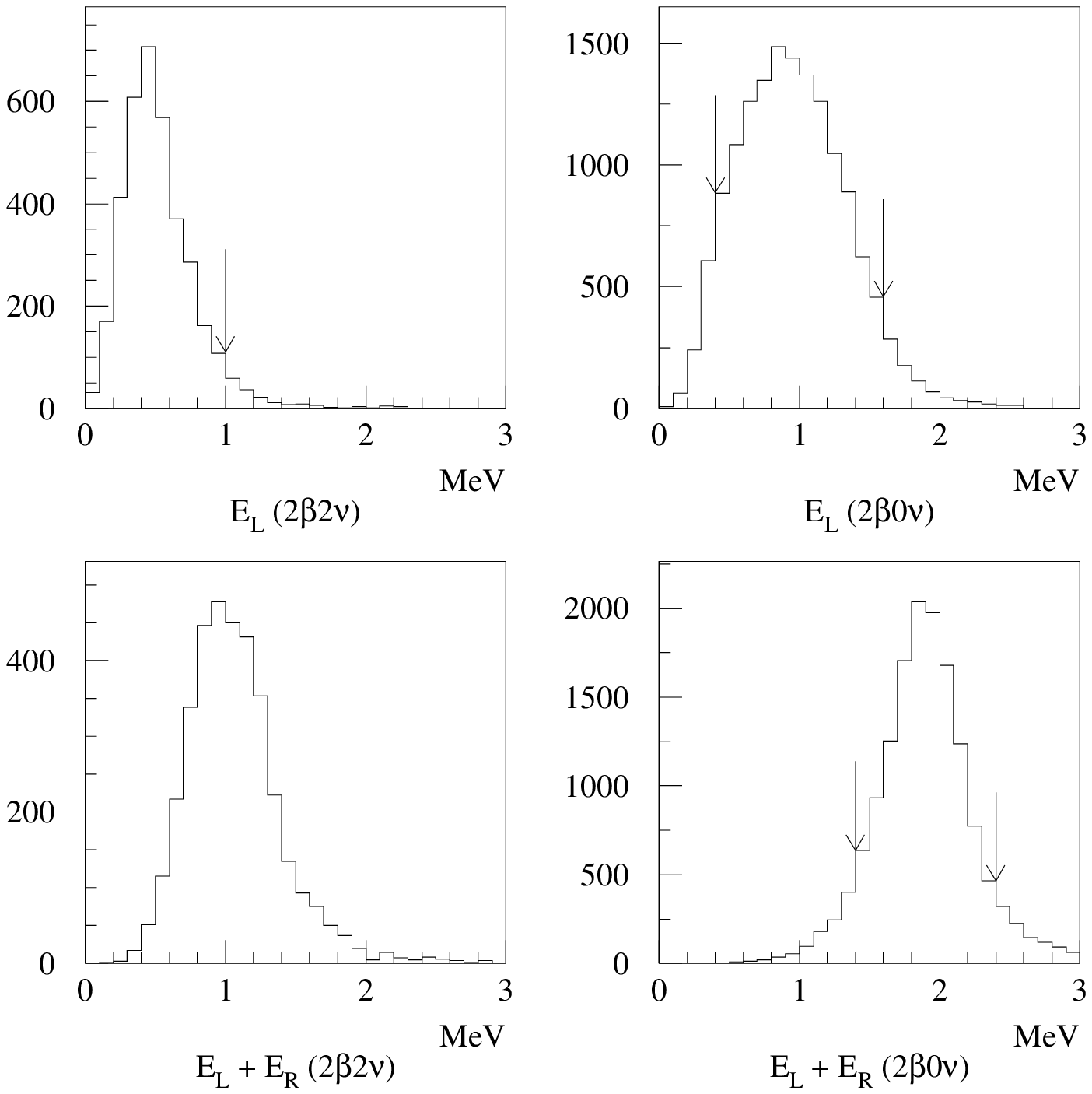}
\caption{Distribution on single electron energy and on sum energy of two electrons for 
Monte-Carlo events}\label{pic_e_mc}
\end{center}\end{figure}

\section{Results}

Distribution along the drift of remaining 909 $2\beta$-events showed substantial 
growth of the number of the events from small drift distances. This 
distribution differed from the same distributions after on-line (see Fig.~\ref{pic_dmexe}) 
and off-line processions. Similar concentration of events was observed 
for the events concerned with $^{214}$Bi decay (see Fig.~\ref{pic_dmexe2}). We assume that these 
events can be concerned with an interaction of the cosmic particles in 
multiwire detectors frames and also with possible concentration growth 
of ions $^{214}$Bi near the detector as a result of diffusion Rn from a 
frame's material. Events from small drift distances with the corresponding 
corrections of efficiency have been thrown out. Number of $2\beta$-candidate 
events was reduced more than twice while efficiency decreased only by 10\%.

Energy spectrum of the single electron for $2\beta$-candidate events 
was more rigid than spectrum of Monte-Carlo events passed full processing. 
Therefore, energy restrictions were chosen according to Monte Carlo 
distributions.

On Fig.~\ref{pic_e_mc} distributions of energy for each electron and total energy of 
two electrons for $2\beta0\nu$ and $2\beta0\nu$ Monte-Carlo events are presented. 

Energy cuts for each electron $E_{L,R} < 1$~MeV for $2\beta2\nu$ decay (see Fig.~\ref{pic_e_mc}) 
were introduced with small efficiency loss. Assuming the same statistics 
to be obtained during an exposition with $^{136}$Xe and all 254 remaining 
events to be background we estimate sensitivity of the detector in 1000 
measurement hours as $T_{1/2}(2\beta2\nu) > 2.7\cdot10^{20}$~years.

To estimate sensitivity for $2\beta0\nu$ decay energy cuts 
$0.4 < E_{L,R} < 1.4$~MeV 
for every electron energy and for the sum energy $1.4 < E_L+E_R < 2.4$~MeV 
were used. Such wide interval for the sum energy was used because the 
electron energy was measured in the methane volumes and losses in Xe 
were not taken into account. Further measurement of ionization losses 
in Xe volume and corresponding narrowing of an interval for the two-electron 
sum energy is planned. With such cuts there were 152 events, that led to 
limit estimation $T_{1/2}(2\beta0\nu) > 5.3\cdot10^{22}$~years for 
1000 measurement hours. 

\section{Conclusion}

In the near future the exposition of TPC detector with $^{136}$Xe will be 
started. Sensitivity of setup can be increased using both the top and 
bottom drift sections. The results discussed were obtained with half of 
detector (only the top drift section). Measurement ionization losses 
will give additional criteria in suppression of background events. 
Including the second half of detector, measurement of ionization losses 
in Xe and increase of measurements time will allow to improve results.

Authors are grateful to M.V.Danilov for constant interest, support and 
discussion of results and possible ways of their improvement. Work is 
supported by the grant of the Russian Basic Research Fond 02-02-16481.



\begin{thebibliography}{99}
\bibitem{1}
S.\,R.\,Elliot, P.\,Vogel, Ann. Rev. Nucl. Part. Sci. \textbf{52}, 115~(2002).
\bibitem{2}
S.\,M.\,Bilenky, hep-ph/0403245.
\bibitem{3}
H.\,V.\,Klapdor-Kleingrothaus, A.\,Dietz, H.\,L.\,Harley, I.\,V.\,Krivosheina, 
Mod. Phys. Lett. \textbf{A16}, 2409~(2001).
\bibitem{4}
H.\,V.\,Klapdor-Kleingrothaus, M.\,Hirsh, Z.\,Phys. \textbf{A359}, 361~(1997).
\bibitem{5}
M.\,Danilov, R.\,De Voe, A.\,G.\,Dolgolenko \emph{et al.}, Phys. Lett. \textbf{B480}, 
12~(2000).
\bibitem{6}
S.\,M.\,Bilenky, S.\,T.\,Petkov, hep-ph/0405237.
\bibitem{7}
C.\,E.\,Aalseth (IGEX Coll.), Nucl. Phys. (Proc. Suppl.) \textbf{B87}, 236~(1999).
\bibitem{8}
M.\,Bakalyarov (Heidelberg-Moscow Coll.), Phys. Rev. \textbf{D55}, 54~(1997).
\bibitem{9}
V.\,Artemiev, E.\,Brakchman, M.\,Ivanovsky \emph{et al.}, Phys. Lett. \textbf{B280}, 
159~(1992).
\bibitem{10}
J.-C.\,Vuilleumier, J.\,Busto, J.\,Farine \emph{et al.}, Phys. Rev. \textbf{D48}, 
1009~(1993).
\bibitem{11}
R.\,Luescher, J.\,Farine, F.\,Boehm \emph{et al.}, Phys. Lett. \textbf{B434}, 
407~(1998).
\bibitem{12}
R.\,Bernabei (DAMA Coll.), Nucl. Phys. (Proc. Suppl.) \textbf{B110}, 88~(2001).
\bibitem{13}
Yu.\,Gavriljuk, A.\,Gangapshov, V.\,Kuz'minov \emph{et al.}, Yadernaya Fizika, 
Intern. Conf. On Non-Accelerator New Physics, (Dubna, 2003).
\bibitem{14}
V.\,A.\,Artemiev, E.\,V.\,Brachman, M.\,A.\,Ivanovsky \emph{et al.}, Yadernaya Fizika 
v.63, 1312~(2000).
\bibitem{15}
V.\,A.\,Artemiev, E.\,V.\,Brachman, M.\,A.\,Ivanovsky \emph{et al.}, NIM A v.477, 414~(2002).
\bibitem{16}
V.\,A.\,Artemiev, E.\,V.\,Brachman, M.\,A.\,Ivanovsky \emph{et al.}, NIM A v.303, 309~(1991).
\bibitem{17}
CERN Program Library (http://cernlib.web.cern.ch/cernlib)
\bibitem{18}
M.\,K.\,Moe and D.\,D.\,Lowenthal, Phys. Rev. \textbf{C22}, 2186~(1980).
\bibitem{19}
LBNL Isotopes Project (http://ie.lbl.gov/databases/databases.html)
\end{thebibliography}
\end{document}